\documentclass{article}
\usepackage{latexsym}
\usepackage{amsfonts}
\usepackage{epsfig}
\usepackage{amssymb}
\usepackage{amsmath}

\def\t{\tau}
\def\be{\begin{equation}}
\def\ee{\end{equation}}
\def\beq{\begin{eqnarray}}
\def\eeq{\end{eqnarray}}
\newcommand{\CC}{\mathbb C}

\newcommand{\R}{\mathbb R}
\newcommand{\Z}{\mathbb Z}

\newcommand{\h}{\,{\bf H}}
\renewcommand{\t}{\,{\bf T}}
\newcommand{\e}{\,{\bf E}}
\newcommand{\f}{\,{\bf F}}
\newcommand{\s}{\,{\bf S}}

\newcommand{\y}{\,{\bf Y}}
 \newcommand{\SL}{\mbox{SL}(2,\mathbb{R})}
 \newcommand{\ls}{{\frak{sl}(2,\mathbb{R})}{}}
\def\lambdabar{{\mathchar'26\mkern-12mu\lambda}}
\def\st#1#2{\underset{#2}{\overset{#1}{\star}}}
\newcommand{\p}{\partial}
\newcommand{\pb}{\bar{\partial}}

\begin{document}

\begin{center}
{\Large BTZ black holes, WZW models and noncommutative geometry\\}

\vspace{.5cm} P.~Bieliavsky${}^{a,}$\footnote{ E-mail :
 bieliavsky@math.ucl.ac.be}, S.~Detournay${}^{b,}$\footnote{ E-mail :
Stephane.Detournay@umh.ac.be, "Chercheur FRIA" },
M.~Rooman${}^{c,}$\footnote{ E-mail : mrooman@ulb.ac.be, FNRS
Research Director} and Ph.~Spindel${}^{b,}$\footnote{ E-mail :
spindel@umh.ac.be}\\

\vspace{.3cm}  {${}^a${\it D\'epartement de Math\'ematique}}\\
{\it   Universit\'e Catholique de Louvain, Chemin du cyclotron,
2,}\\
{\it 1348 Louvain-La-Neuve, Belgium}\\
\vspace{.2cm} {${}^b${\it M\'ecanique et Gravitation}}\\ {\it
Universit\'e de Mons-Hainaut, 20
Place du Parc}\\
{\it 7000 Mons, Belgium}\\
\vspace{.2cm}  {${}^c${\it Service de Physique th\'eorique}}\\
{\it Universit\'e Libre de Bruxelles, Campus Plaine, C.P.225}\\
{\it Boulevard du Triomphe,
B-1050 Bruxelles, Belgium}\\
\end{center}
\vspace{.1cm}

\begin{abstract} This note is based on a talk given by one of the
authors at the "Rencontres Math\'ematiques de Glanon", held in
Glanon in July 2004. We will first introduce the BTZ black hole,
solution of Einstein's gravity in 2+1 dimensions, and emphasize
some remarkable properties of its geometry. We will essentially
pay attention to the non-rotating black hole, whose structure is
significantly different to the generic case. We will then turn the
some aspects of string theory, namely the emergence of
non-commutative geometry and the embedding of the BTZ black hole
as an exact string background using the Wess-Zumino-Witten (WZW)
model. We will show the existence of winding symmetric WZW
D1-branes in this space-time from the geometrical properties of
the non-rotating black hole. Finally, we will introduce strict
deformations of these spaces, yielding an example of
non-commutative lorentzian non-compact space, with non-trivial
causal structure.
\end{abstract}

\section{Geometry of the BTZ black holes}
Vacuum Einstein's equations in 2+1 dimensions with negative
cosmological constant admit black hole solutions, as first quoted
in \cite{DT}. A remarkable property of these solutions lies in the
fact that they arise from identifications of points of $AdS_3$ by
a discrete subgroup of its isometry group, as shown by Ba\~nados,
Henneaux, Teitelboim and Zanelli \cite{BTZ,BHTZ}. They are
therefore usually referred to as {\itshape BTZ black holes} in the
physics literature. According to the type of subgroup, the vacuum,
extremal or generic (rotating or non-rotating) massive black holes
are obtained.
\subsection{Three-dimensional anti-de Sitter space}
The three dimensional anti de-Sitter space ($AdS_3$) is the
maximally symmetric solution
of vacuum
Einstein's equation in 2+1 dimensions with negative cosmological
constant $\Lambda$. It is defined as the hyperboloid
\begin{equation}\label{hyperb}
  u^2 + t^2 - x^2 - y^2 = -1/\Lambda \quad,
  \end{equation}
  embedded in the four dimensional flat space with metric $ds^2 = -du^2 - dt^2 + dx^2 +
  dy^2$. The Ricci tensor of this solution satisfies $R_{\mu\nu} =
  2\Lambda g_{\mu\nu}$ and the scalar curvature is $R=6\Lambda$.
  Hereafter, we will set $\Lambda=-1$.

  It follows from (\ref{hyperb}) that $AdS_3$ can equivalently by
  defined as the (universal covering of) the simple Lie group
  $\SL$ :
  \begin{equation}\label{AdSl}
 AdS_3 \cong \SL = \{g = \left(
         \begin{array}{cc} u+x & y+t\\ y-t & u-x
         \end{array}
         \right)| x,y,u,t\in\mathbb{R},\ \det g =1 \} \qquad .
         \end{equation}
We endow $\SL$ with its bi-invariant Killing metric, defined at
the identity $e$ using the Killing form $B$ of its Lie algebra
$sl(2,\R)$, identified with the tangent space at the identity: \be
 \beta_e (X,Y) = B(X,Y) = \frac{1}{2}\mbox{Tr}(X.Y) \quad, X,Y \in
 sl(2,\R) \quad ,
\ee

where the last equality holds because we are dealing with a matrix
group, the trace being understood as the usual matricial one (i.e.
in the fundamental representation). It can be checked, given a
parametrization $g(X^\mu)$ of the group manifold , that the
invariant metric $\mbox{Tr}(g^{-1} dg g^{-1} dg)$ actually
coincides with the $AdS_3$ metric (see sect.\ref{ConfStr}). The
Lie algebra
 \begin{equation}
         \label{matsl} sl(2,\R)=\{\left(\begin{array}{cc} z^H & z^E \\ z^F &
         -z^H
         \end{array}\right):= z^H\h+z^E\e+z^F\f\} \qquad ,
         \end{equation} is expressed in terms of the generators $\{\h , \e, \f
         \}$ satisfying the commutation relations:
         \begin{equation}
         \label{comrel} [\h ,\e]=2\e \quad , \quad [\h,\f]=-2\f \quad ,
         \quad [\e,\f]=\h \quad .
         \end{equation}
For later purpose, we also define the one-parameter subgroups of
$\SL$ :
 \begin{equation} A=\exp(\R\,\h)\ ,\quad N=\exp(\R\,\e)\
         ,\quad K=\exp(\R\,\t)\ ,\ \qquad ,
\end{equation}
which are the building blocks of its Iwasawa decomposition (see
\cite{Barut,Helgason}) :

\be \label{Iwasawa}
 K \times A \times N \rightarrow \SL(2,\R) : (k,a,n)
\rightarrow kan \,\, \mbox{or} \, \, ank \, ,
  \,k\in K, \, a\in A, \, n\in N \quad,
 \ee
the mapping being an analytic diffeomorphism of the product
manifold $K \times A \times N $ onto Sl$(2,\R)$ (see
\cite{Helgason}, p.234)

\subsubsection{Isometry group of $AdS_3$}\label{Isom}
By construction, the isometry group of $AdS_3$, denoted by
$Iso(AdS_3)$, is the four-dimensional Lorentz group $O(2,2)$ (see
(\ref{hyperb})). It is constituted by four connected parts, whose
identity component is $SO(2,2)$. The other components can always
be reached using a parity ($x\rightarrow -x$) and/or time reversal
($t\rightarrow -t$) transformation.
In terms of the coordinates of (\ref{hyperb}), the Killing vector
fields are expressed as $J_{ij} = x_j
\partial_i - x_i \partial_j$, where $x^i = (u,t,x,y)$.
 Now, because of (\ref{AdSl}) and due to the bi-invariance of the
 Killing metric, $SO(2,2)$ is locally isomorphic to
 $\SL \times \SL$ (they share the same Lie algebra) , the action being given by
 \be\label{actisom}
  (\SL \times  \SL) \times  \SL \rightarrow  \SL :
  ((g_L,g_R),z) \rightarrow g_L \, z \, g_R^{-1} \quad.
  \ee
This action clearly corresponds to the component of $Iso(AdS_3)$
connected to the identity transformation (because $\SL$ is
connected). The full action of $Iso(AdS_3)$ (including the
components not connected to the identity) can be obtained by
considering the parity and time reversal transformations
\begin{eqnarray}
{\cal P}(g)&=&\left
(\begin{array}{cc}u-x&y+t\\y-t&u+x\end{array}\right)\quad,\label{revx}\\
    {\cal T}(g)&=&\left
(\begin{array}{cc}u+x&y-t\\y+t&u-x\end{array}\right)\quad.\label{revt}
\end{eqnarray}

There exists a Lie algebra isomorphism between $sl(2,\R) \times
sl(2,\R)$ and $iso(AdS_3) \equiv Lie(Iso(AdS_3)) $. It is given by
 \be\label{Iso}
 \Phi : sl(2,\R) \times sl(2,\R) \rightarrow iso(AdS_3) : (X,Y)
 \rightarrow  \overline{X} - \underline{Y} \quad ,
\ee

where $\overline{X}$ (resp. $\underline{Y}$) denotes the
right-invariant (resp. left-invariant) vector field on $\SL$
associated to the element $X$ (resp. $Y$) of its Lie algebra, that
is \beq
 \overline{X}_g &=& \frac{d}{dt}_{|0} L_{\exp t X} \, g =
 \frac{d}{dt}_{|0} \exp (t X) \, g = X g \quad, \\
\underline{Y}_g &=& \frac{d}{dt}_{|0} R_{\exp t X} \, g =
 \frac{d}{dt}_{|0}  g\, \exp (t Y) = g Y \quad,
\eeq where $L_g$ and $R_g$ denote the left and right translations
on the group manifold. The two last equalities hold because $\SL$
is a matrix group.

Thus, to the generator $(X,Y) \in sl(2,\R)\times sl(2,\R)$, we
associate the one-parameter subgroup of (the connected part of)
$Iso(AdS_3)$
\begin{equation}
 \Psi_t(g)= \exp(tX) \, g \, \exp(-tY) \quad,\quad g\in \SL, t\in
 \R \quad.
 \end{equation}
The classification, up to conjugation, of the one-parameter
subgroups of $\SL \times \SL$ has been achieved in
\cite{BHTZ,BRS}. Two subgroups $(G_L,G_R)$ and $(G'_L,G'_R)$  are
said to be conjugated if

{\itshape (i)} they can be related by combinations of
transformations $\cal P$ and $\cal T$;

{\itshape (ii)} there exists $g_L$ and $g_R$ $\in \SL$ such that
$G_i = g_i \, G'_i\, g_i^{-1}$, for $i=L,R$.
Note that {\itshape (ii)} corresponds to a conjugation by an
element of $SO(2,2)$, while {\itshape (i)} corresponds to an
element not connected to the identity. The Killing vectors of
$AdS_3$ (i.e. the generators of the one-parameter subgroups of
$\SL \times \SL$) are then all of the form
 \be
 (a X,b Y), \quad a,b \in \R \quad, \quad X,Y = \{\h,\t,\e,\f\} \quad.
 \ee

\subsubsection{Conformal structure/representation}\label{ConfStr}
A system of coordinates covering the whole of the $AdS_3$ manifold
(up to trivial polar singularities) may be introduced by setting

\beq u= l \cosh\xi\,\cos\lambda\qquad ,& \hspace{10mm} &
t= l \cosh\xi\,\sin\lambda \qquad ,\nonumber\\
x=l \sinh\xi\,\cos\varphi \qquad ,& \hspace{10mm} & y=l
\sinh\xi\,\sin\varphi \quad, \eeq

in which the invariant metric on $AdS_3$ reads
\begin{equation}
\label{metric} ds^2= l^2( -\cosh^2\xi\, d\lambda^2 + d\xi^2 +
\sinh^2\xi\,d\varphi^2)\qquad,
\end{equation}
with $l^2 = -\frac{1}{\Lambda}$. The further change of radial
coordinate:
\begin{equation}
\tanh\xi=\sin r \hspace{10mm} r \in[0,\pi/2[ \qquad ,
\end{equation}
yields a conformal embedding of $AdS_3$ in the Einstein static
space \cite{HE}:
\begin{equation}\label{he} ds^2=\frac{l^2}{\cos^2 r }\left( -d\lambda^2+d r
^2+\sin^2 r \,d \varphi ^2\right) \qquad .
\end{equation}

\subsection{BTZ black holes}
\subsubsection{General construction}
Einstein gravity with negative cosmological constant in 2+1
dimensions admits axially symmetric and static black hole
solutions, characterized by their mass $M$ and angular momentum
$J$. One distinguishes between the extremal ($|J|=M l$), massless
($J=M=0$), non-rotating massive ($J=0$,$\,M>0$) and generic
($0<|J| < M l$) black holes. These solutions share many common
properties with their 3+1 dimensional counterparts. Ba\~nados,
Henneaux, Teitelboim and Zanelli observed \cite{BHTZ} that all of
these solutions could be obtained by performing identifications in
$AdS_3$ space by means of a discrete subgroup of its isometry
group, which we described in section \ref{Isom}. According to the
choice of Killing vector $\Xi = (X_L,X_R)$, the different black
holes can be obtained by the quotient procedure. The mass and the
angular momentum of the solution will be related to $\Xi$ via \beq
M(\Xi)&=& \frac 1 2 \left ( \|X_L\|^2+\|X_R\|^2 \right )\qquad,\label{mass}\\
J(\Xi)&=&\frac 1 2  \left ( \|X_L\|^2-\|X_R\|^2 \right
)\qquad,\label{angm} \eeq
 where $\|X\|^2$ is the square of the Killing norm
$\beta_e (X,X)$\footnote{Note that in \cite{BHTZ}, $M$ and $J$
associated to a Killing vector $\Xi = \frac{1}{2}
\omega^{ab}J_{ab}$ are given by $I_1 = \omega_{ab}\omega^{ab} =
-2M$ and $I_2 = \frac{1}{2}\epsilon_{abcd}\omega^{ab}\omega^{cd} =
-2\frac{|J|}{l}$. This coincides with definitions (\ref{mass}) and
(\ref{angm}).}.

In the following, we will focus on the non-rotating massive black
hole. In this situation, because of (\ref{mass}) and (\ref{angm}),
$X_L$ and $X_R$ must belong to the same space-like adjoint
orbit\footnote{Indeed, for $J$ to vanish, the left and right
generators must belong to the same orbit. Then, the condition $M >
0$ excludes the time and light-like orbits. Thus, we are only left
with the category $I_b$ in the classification p6 of \cite{BRS}}.

In this case, we define the {\itshape BHTZ subgroup} as the
one-parameter subgroup of $\SL \times \SL$ generated by the
Killing vector $\Xi = a(\overline{\h} - \underline{\h})
=(a\h,a\h)$\footnote{with a slight abuse of notation : actually,
$\frac{d \psi_t (z)}{dt} = \h \psi_t(z) - \psi_t(z) \h =
\overline{\h}_{\psi_t(z)} - \underline{\h}_{\psi_t(z)} =
\Xi_{\psi_t(z)}$} : \be \label{BHTZ}
 \psi_t (z) = \exp(t a \h) \, z \, \exp(-t a \h) \, , \quad z\in
 \SL.
 \ee
A non-rotating massive BTZ black hole is then obtained as the
quotient $\SL / \psi_\Z$ by an isometric action of $\Z$, that is,
by performing the following identifications :

 \be \label{Identif}
 z \sim \exp(n a \h) \, z \, \exp(-n a \h) \quad,\quad n\in \Z \quad.
\ee

The resulting space-time displays a black hole structure
\cite{BHTZ}.
Namely, it exhibits {\itshape horizons}, i.e. light-like surfaces
separating two regions, one of which can be causally connected to
the space-like infinity (the exterior region), and another one
causally disconnected from it (the interior region). Inside the
horizon, any light-ray necessarily ends on the singularity.

Note that, to avoid closed time-like curves
in the quotient space, the regions of $AdS_3$ where the orbits of
the BHTZ action (\ref{BHTZ}) are time-like ,or equivalently where
$\beta_z (\Xi,\Xi) < 0$, must be excluded. We will henceforth
consider an open and connected domain ${\cal U}\subseteq AdS_3$
where $\beta_z (a \h,a \h) > 0 \quad, \forall z\in {\cal U}$.
 The black holes singularities ${\cal S}$ are defined as the
 surfaces where the identifications becomes light-like :
 \be
 {\cal S} =\{ z \in AdS_3 \,\, | \,\,  \beta_z(\Xi,\Xi) = 0\} \quad.
\ee

\subsubsection{Massless non-rotating BTZ as twisted bi-quotient
space/Global structure}

We would now like to focus on the global geometry of the
non-rotating massive BTZ black hole \cite{BRS}. The aim is
essentially to reduce ourselves to a two-dimensional problem by
finding a foliation of $AdS_3$ whose leaves are stable under the
BHTZ action, and are all isomorphic (so that we can restrict
ourselves to the leaf through the identity). Let us first
introduce an external automorphism of $\ls$, which can be chosen
as
\begin{equation}
\label{defsig} \sigma(\h)=\h,\ \sigma(\e)= -\e ,\ \sigma(\f) = -\f
\qquad .
\end{equation}
Viewing $\SL$ as a subgroup of ${GL(2,\mathbb{R})}$, one may
express the automorphism $\sigma$ as:
\begin{equation}
\sigma=Ad(\h) \qquad \mbox{where} \qquad \h=\left(
\begin{array}{cc}
1 & 0\\
0 & -1
\end{array}
\right)
  \in GL(2,\mathbb{R}) \qquad .
\end{equation}
The corresponding external automorphism of $\SL$ (which we again
denote by $\sigma$) is $\sigma (g) = \h \,g \h$ (note that $\h =
\h^{-1}$). We then consider the following {\itshape twisted}
action of $\SL \subset SO(2,2)$ on itself :
 \be \label{TwistedAction}
 \SL \times \SL \rightarrow \SL: (g,z) \rightarrow \tau_g(z) = g \, z\,
 \sigma(g^{-1}) \quad.
\ee The fundamental vector field associated to this action is
given by : \be X^*_z =  \frac{d}{dt}_{|t=0} \tau_{\exp(-tX)}(z) =
-X z + z \sigma(X)
 \quad,
 \ee
 by noting that $\sigma(\exp (tX))= \exp (t \, \sigma(X))$, with
 an evident
 abuse of notation.
We note that the BHTZ action $\Z \times \SL$ ( eq.(\ref{BHTZ}))
can equivalently be written as \be \label{BHTZTwisted}
 \psi_t (z) = \tau_{\exp (a\,t\,\h)} (z) \quad .
 \ee
The orbit of a point $z$ under this action is \be \label{orbit}
 {\cal O}_z =  \{\,\, \tau_g (z) \quad, \,
 g\in\SL\} \quad
 \ee
and is usually referred to as a {\itshape twisted conjugacy
class}. This particular structure will become important in section
2 where we will discuss D-branes in Wess-Zumino-Witten models.
 Using the matrix representation of (\ref{AdSl}), we find that
 ${\cal O}_z$ is constituted by the elements $g\in\SL$ such that
 $x(g) = x(z)=cst$. We see from (\ref{hyperb}) that the orbits
 are two-dimensional one-sheet hyperboloids in flat
 four-dimensional space. We won't make use of this representation,
 but adopt a more convenient one.

 The global description of the black hole relies on the following
 proposition. Consider the application :
 \be \label{twistedI}
 \phi : K \times A \times N \rightarrow \SL : (k,a,n) \rightarrow
 \phi(k,a,n) = \tau_{kn}(a) \quad.
 \ee
 This application is a global diffeomorphism.
Furthermore, we observe, by letting $g=k n a' \in \SL$ that \be
  \tau_g (a) = \phi(a,n,k) \quad \forall a \in A. \ee
Therefore, the orbit of $a \in A$ under the twisted bi-action is
given by \be {\cal O}_a = \phi(a,N,K) \cong NK = G/A \quad , \ee
where the last equality holds because of the Iwasawa decomposition
of $G=\SL$ (\ref{Iwasawa}). The application can be rewritten as

\be \label{phibis} \phi : A \times G/A \rightarrow G :
(a,[g])\rightarrow g \, a \, \sigma(g^{-1}) \quad , \ee

where $[g]  = \{ gA,\, g \in G\}\in G/A.$ This is exactly the
decomposition we needed to reduce the geometrical description of
$AdS_3$ and the black hole from three to two dimensions. Indeed,
it provides us with a foliation (a trivial fibration) of $AdS_3$
over $A \simeq \R$ whose leaves are two-dimensional orbits of the
action (\ref{TwistedAction}). Each of the leaves is stable under
the BHTZ subgroup, that is, $ z \in {\cal O}_a \Rightarrow \psi_t
(z) \in {\cal O}_a ,\, a\in A\,,z\in \SL$. Moreover, the BHTZ
action is fiberwise\footnote{i.e., we may consider the action of
the BHTZ subgroup "leaf by leaf".} , which means that

\be \label{fiberwise}
 \tau_h (\phi(a,[g])) = \phi(a,h.[g]) = \phi(a,[hg]) \quad.
 \ee

This, with the fact that the orbits are all diffeomorphic,
allow us to restrict ourselves to the study of the leaf through
the identity and focus on the space $G/A$\footnote{The stabilizer
of $e$ for the twisted bi-action is $Stab_G(e)=A$, $G$ acting on
${\cal O}_e$. Therefore, ${\cal O}_e$ is homeomorphic to $G/A$.
This is another way of seeing the orbits as homogeneous spaces.
Last, to go to a general orbit, we use the fact that all the
orbits are diffeomorphic.}. This space can be realized as the
adjoint orbit of $\h$ in $\ls$. Indeed, consider the application
\be
 \Upsilon : G/A \rightarrow Ad(G)\h : [g] \rightarrow \Upsilon([g]) =
 Ad(g)\h\quad.
\ee Because $Ker \Upsilon = \{[e],[-e]\}$, $G/A$ is a $\Z_2$
covering of $Ad(G)\h$. It is well defined because $Ad(a)\h = \h \,
\forall a\in A$.

With $A\cong\R$ and using the application $\Upsilon$, $\phi$ can
further be rewritten as

\be \label{diffeoGl2} \phi : \R \times Ad(G)\h \rightarrow G :
(\rho,Ad(kn)\h) \rightarrow \phi(\rho,Ad(kn)\h) =
\tau_{kn}(\exp(\rho\h)) \quad. \ee

 Restricting $\phi$ to the leaf at identity ${\cal O}_e$ (for
which $\rho=0$) , we define the application

\be \label{iota} \iota : Ad(G)\h \rightarrow {\cal O}_e : Ad(kn)\h
\rightarrow \tau_{kn} (e) \quad, \ee and consider the adjoint
orbit of $\h$, which corresponds to a one-sheet hyperboloid in
$\ls$

The orbit of  a point $X$ in the hyperboloid (i.e. corresponding
to a point $x$ in the leaf at the identity) under the BHTZ
subgroup are the intersections of the planes perpendicular to the
$\h$-axis, passing through $X$. Indeed, let $X= Ad(kn)\h$,
corresponding to the point $x=\tau_{kn}(e) \in {\cal O}_e$. The
orbit of $x$ under the BHTZ subgroup is \beq \psi_{n}(x) =
\tau_{\exp(n \h a)}(x) = \tau_{\exp(n \h a)kn}(e) \quad, \eeq
corresponding to the curve $Ad(\exp (n \h a)) X$ on $Ad(G)\h$. One
then checks,
 using the $Ad$-invariance of the Killing form, that
\be \beta_e (\h , Ad(\exp(n \h a)X - X) = 0 \quad. \ee

Let us summarize. To obtain the non-rotating massive BTZ black
hole space-time, we have to perform discrete identifications in
$AdS_3$ along the orbits of the BHTZ action, or equivalently of a
twisted bi-action (see eq.(\ref{BHTZTwisted})). We know from
(\ref{fiberwise}) that the identifications can be performed "leaf
by leaf", all the leaves being diffeomorphic. We also know that,
in order to avoid closed time-like curves in the quotient space,
we have to restrict ourselves to an open and connected domain
${\cal U}\subseteq AdS_3$ where $\beta_z (\Xi,\Xi) > 0 \quad,
\forall z\in {\cal U}$. Therefore we have to identify the
corresponding region in $Ad(G)\h$. It can be seen that the region
where the orbits of the BHTZ action are space-like corresponds to
\be \label{BonDomaine2}
 \{ X = x^{\h} \h + x^{\e} \e + x^{\f} \f \in Ad(G)\h \quad  | \quad -1< x^{\h} < 1
 \} \quad.
 \ee

\begin{figure}[hbt]
\begin{center}

{\epsfig{file=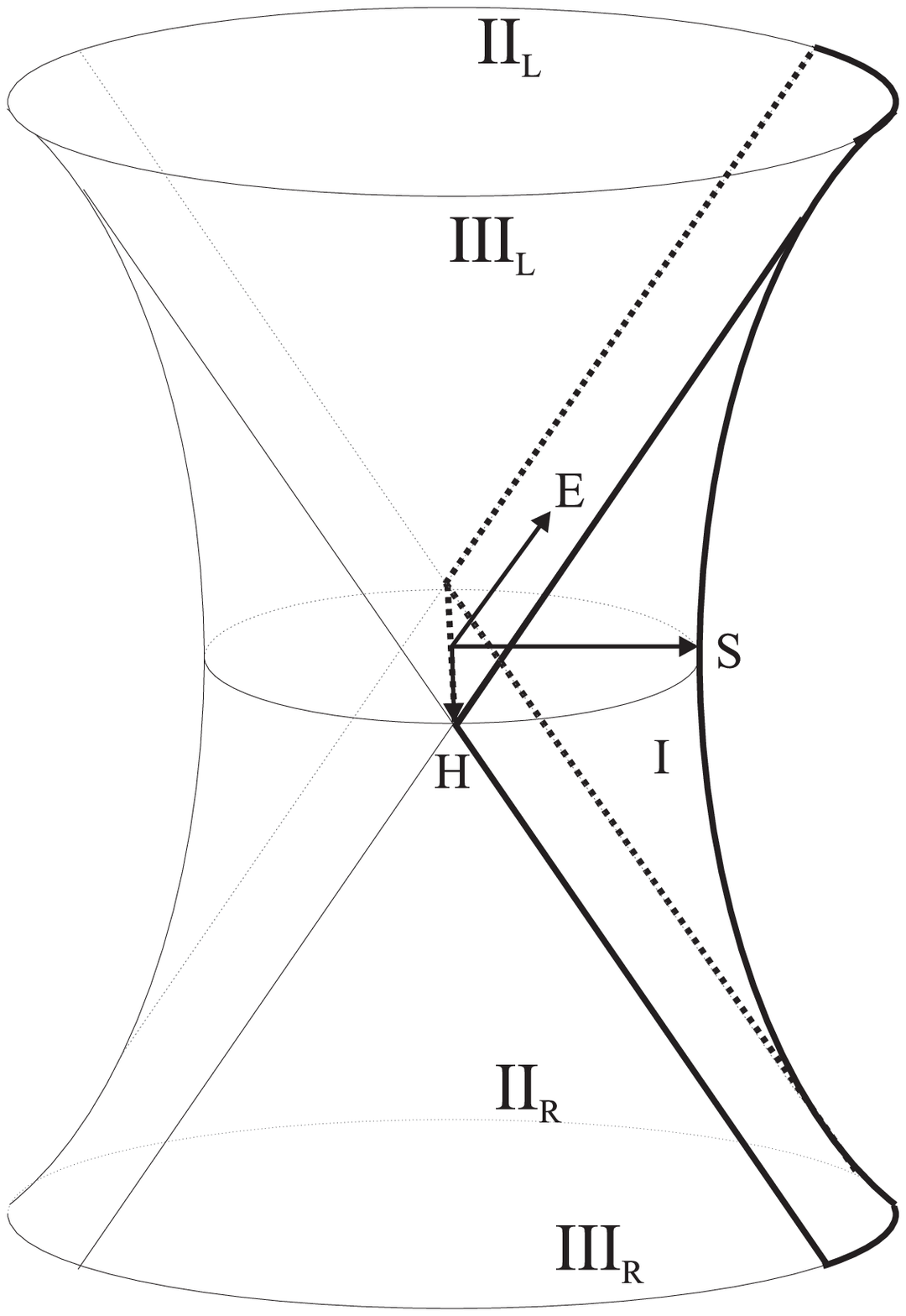,scale=0.7}}
\caption[]{\label{OrbiteBD} An adjoint orbit of $\h$ under $\SL$,
identified to a leaf of the foliation.The region $I$ corresponds
to one region where the BHTZ orbits are space-like.}

{\epsfig{scale=0.7,file=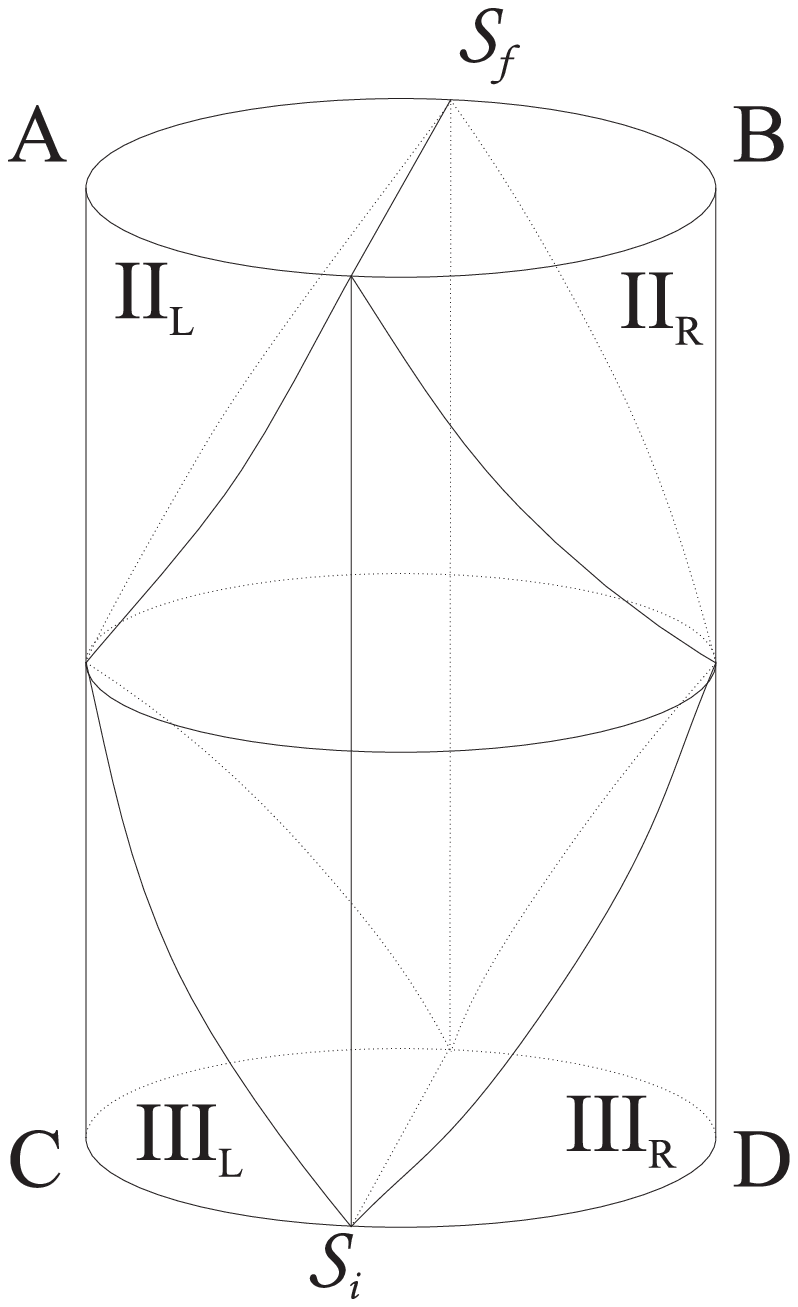}}
\caption[]{\label{Cylindre1} A conformal picture of $AdS_3$, seen
as an Einstein cylinder (see eq. (\ref{he}) ). A region where
$\Xi$ is space-like (a black hole region) is represented between
four light-like surfaces.}

\end{center}
\end{figure}

One of these regions in represented in figure \ref{OrbiteBD} and
labelled $I$. It can be parameterized as\footnote{Note that a
parametrization of the whole hyperboloid would be given by
 $Ad(\exp(\beta/2 \t)\exp(\gamma/2
\h)) \s $, leading to a global metric on $AdS_3$.}

\be
 X = Ad\left(\exp(\frac{\theta}{2}\h) \, \exp(-\frac{\tau}{2}
 \t)\right)\, \h \quad , \, 0<\tau<\pi \, ,\,  -\infty<\theta<+\infty
 .
\ee

In this parameterization, the orbits of the BHTZ action are the
lines $\tau=cst.$. Using (\ref{iota}), any point $x\in {\cal O}_e$
belonging to ${\cal U}$ can be parameterized by the two
coordinates $(\theta,\tau)$ via

\be \label{CoordOe}
 z(\theta,\tau) = \tau_{\exp(\frac{\theta}{2})\h \,
 \exp(-\frac{\tau}{2})
 \t} (e)
 \ee

 Note that $e$ and $-e$ belong to the boundary of ${\cal U}$.
 Finally, we obtain a global parameterization of points in ${\cal
 U}$ with the help of (\ref{phibis}), as

\be \label{CoordGlob} z(\rho,\theta,\tau) =
\tau_{\exp(\frac{\theta}{2}\h) \,
 \exp(-\frac{\tau}{2}
 \t)}(\exp(\rho\h)) \quad .
 \ee
The action of the BHTZ reads in these coordinates

\be (\tau , \rho, \theta) \rightarrow (\tau, \rho, \theta + 2 n a)
\quad . \ee

Eq. (\ref{CoordGlob}) allows to derive a global expression for the
metric of the black hole :
\be \label{Metric} ds^2= d\rho^2 + \cosh^2\rho(-d\tau^2 +
\sin^2\tau d\theta^2) \quad, \quad -\infty < \rho < +\infty,\,
0<\tau<\pi,\, 0<\theta<2a \quad. \ee The black hole has topology
$S^1 \times \R^2$.

Let us turn to some pictures displaying the global structure of
the black hole. Recall from section \ref{ConfStr} that $AdS_3$ is
conformal to the three-dimensional Einstein static space. The time
axis $\lambda$ (see eq.(\ref{he})). Fig.\ref{Cylindre1}, displays
the region ${\cal U}$ where the Killing vector $\Xi$ is
space-like, which is comprise between the four light-like
surfaces.

\begin{figure}[hbt]
\begin{center}

{\epsfig{scale=0.7,file=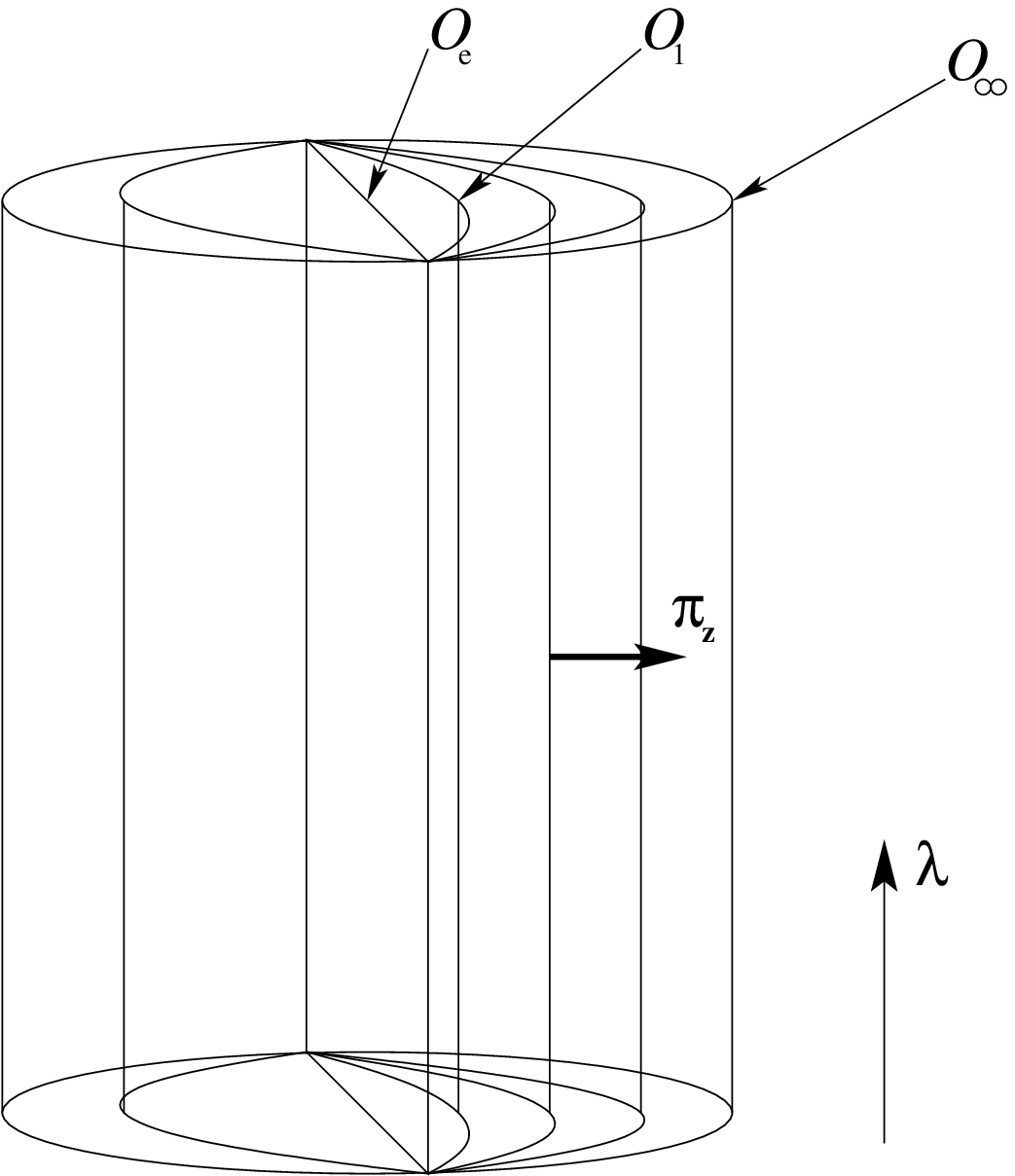}}
\caption[]{\label{foliation}Foliation of $AdS_3$ in twisted
conjugacy classes ($\rho=cst$ surfaces).}

{\epsfig{scale=0.7,file=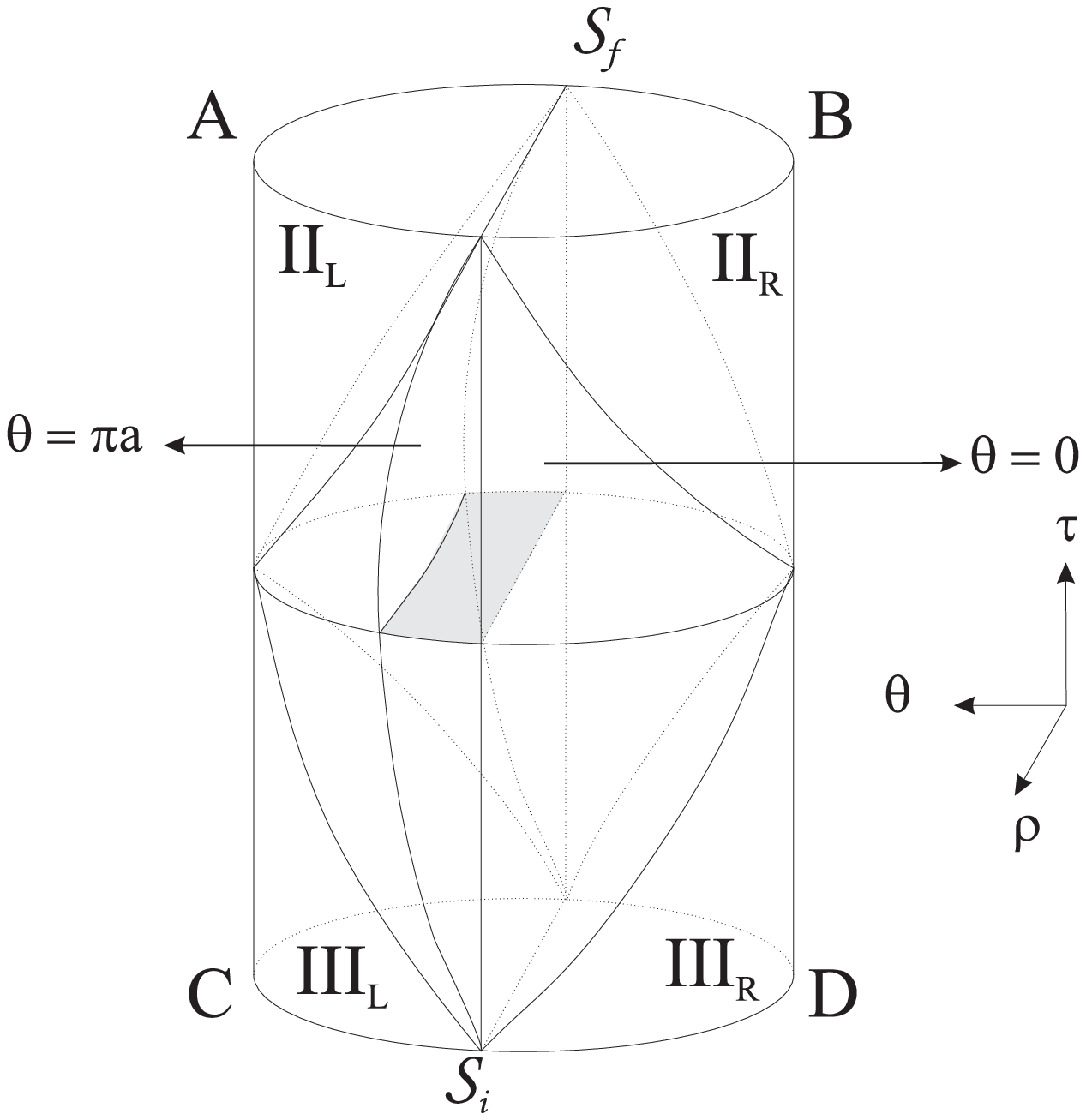}}
\caption[]{\label{BTZCyl}Dynamical evolution of the black hole,
from its initial singularity ${\cal S}_i$ to the final singularity
${\cal S}_f$. The shaded region represents a timelike section of a
fundamental domain of the BHTZ action. }

\end{center}
\end{figure}

The next one (Fig.\ref{foliation}) represents the foliation of
$AdS_3$ by the two-dimensional space-like $\rho=cst.$ leaves.

Figure \ref{BTZCyl} represents the dynamical evolution of the
black hole, from its initial to its final singularity (denoted
${\cal S}_i$ and ${\cal S}_i$) , using the global coordinates
$(\tau,\rho,\theta)$. The shaded region is a fundamental domain of
the BHTZ action.

\begin{figure}[hbt]
\begin{center}

{\epsfig{scale=0.7,file=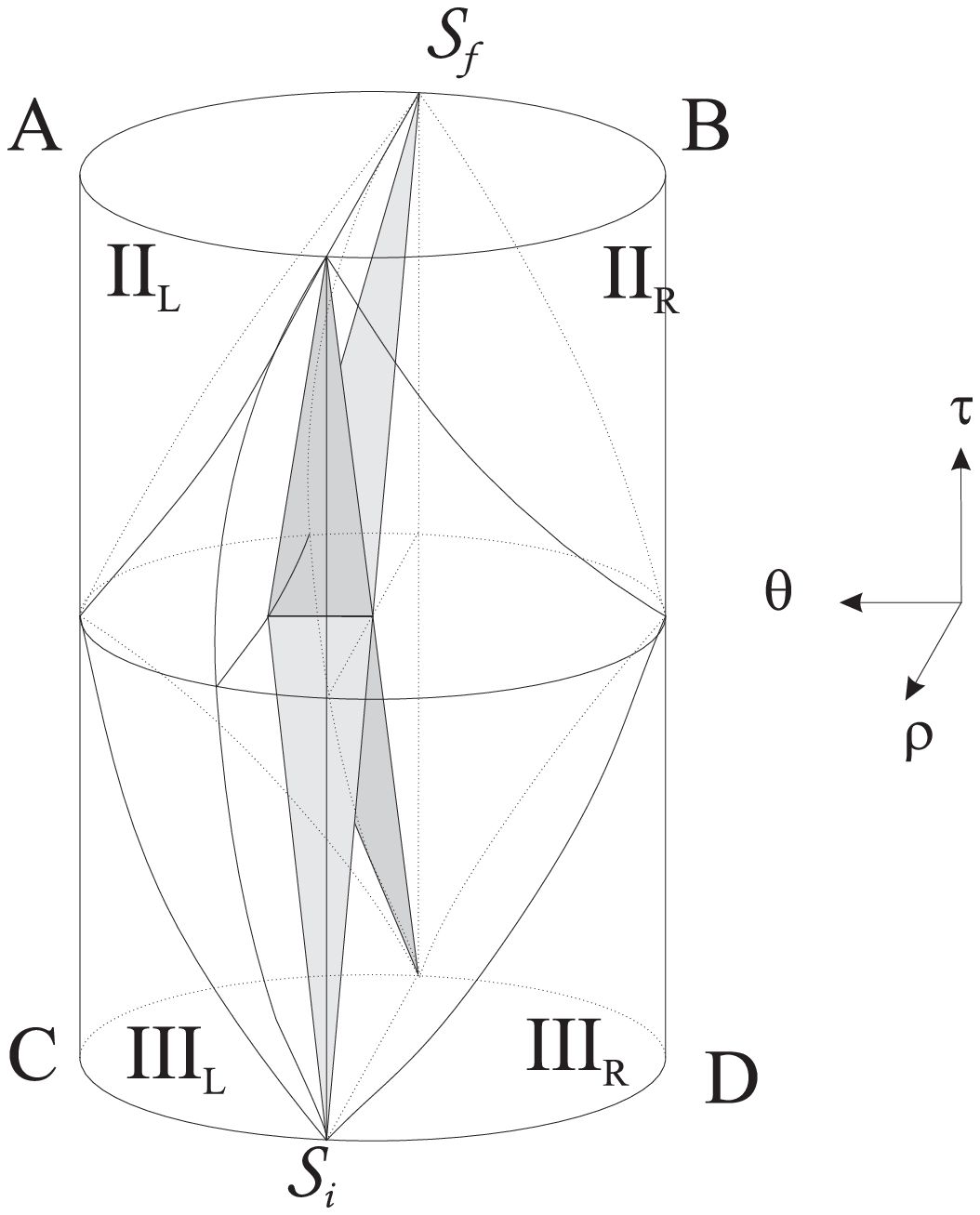}}
\caption[]{\label{BTZHor} Representation of the black hole's
horizons. These are light-like surfaces separating two regions,
one of which can be causally connected to the space-like infinity
(the exterior region), and another one causally disconnected from
it (the interior region).}

{\epsfig{scale=0.7,file=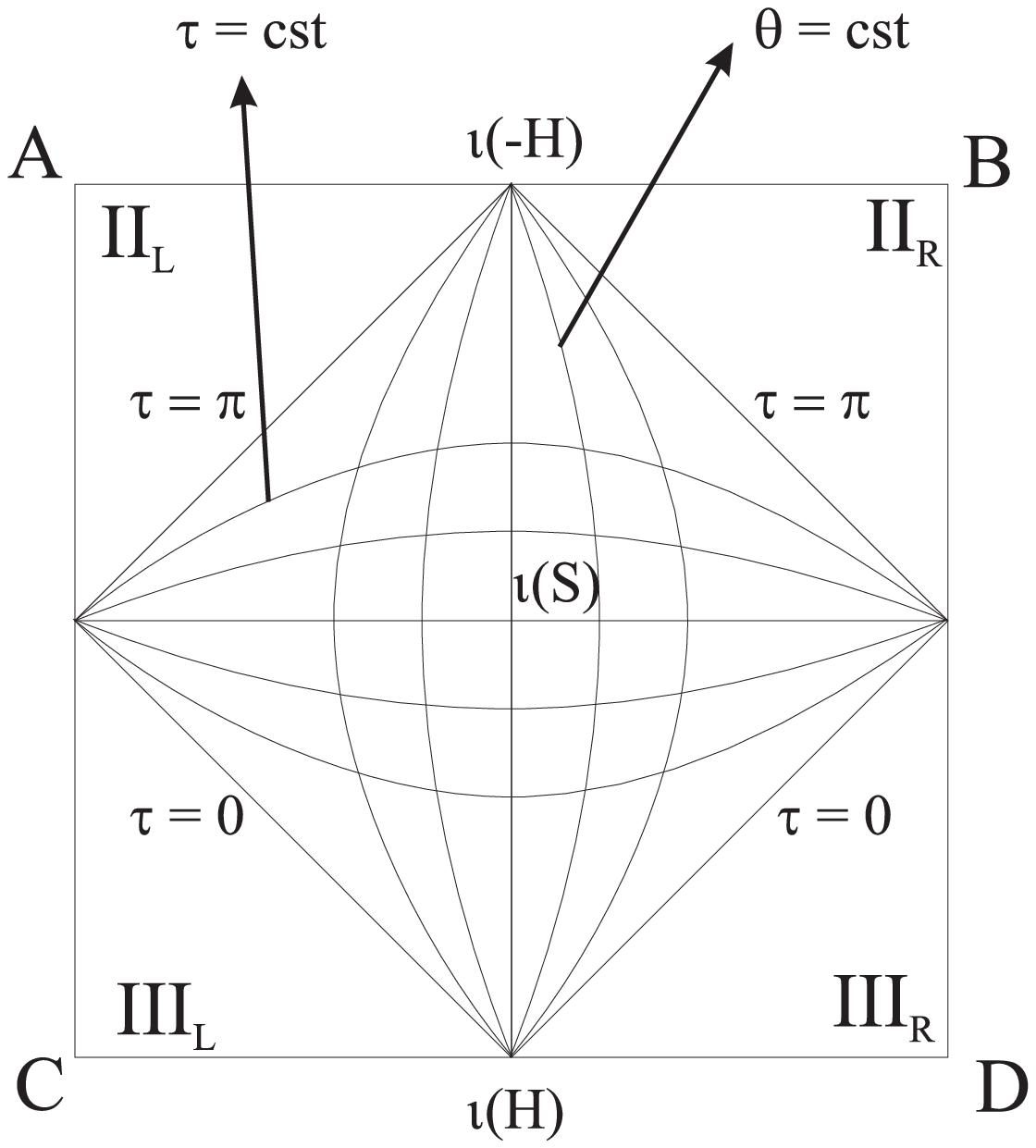}}
\caption[]{\label{feuillerho} A $\rho=0$ section of
Fig.\ref{Cylindre1}, where the coordinate lines are represented.
The orbits of the BHTZ subgroup correspond to $\tau=cst$. }

\end{center}
\end{figure}

In Fig. \ref{BTZHor}, the black hole horizons are drawn.

From these three-dimensional pictures, we can also go to two
different two-dimensional visions. Fig.(\ref{feuillerho}) is a
section of Fig.(\ref{Cylindre1}) by the $\rho=0$ surface. We
represented the coordinate lines $\theta=cst.$ and $\tau=cst.$
(the latter corresponding to orbits of the BHTZ subgroup). On the
straight lines, the identifications become light-like. Note that,
after identifications, the manifold structure is destroyed at
$e=\iota(\h)$ and $-e=\iota(-\h)$, which are fixed points of the
BHTZ action (see sect.$5.8$ of \cite{HE}, and \cite{BHTZ}).
Everywhere else in $I$, the action of the BHTZ subgroup is
properly discontinuous.

\begin{figure}[hbt]
\begin{center}

{\epsfig{scale=0.7,file=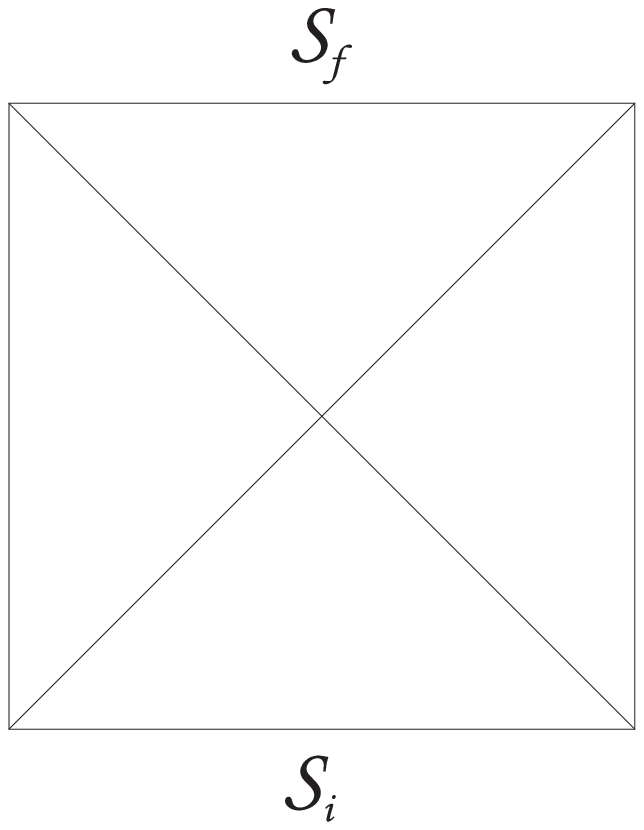}}
\caption[]{\label{Penrose2D}A $\theta=0$ section of
Fig.\ref{Cylindre1}, yielding a Penrose diagram of the black hole.
The oblique lines represent the black hole's horizons.}

{\epsfig{scale=0.7,file=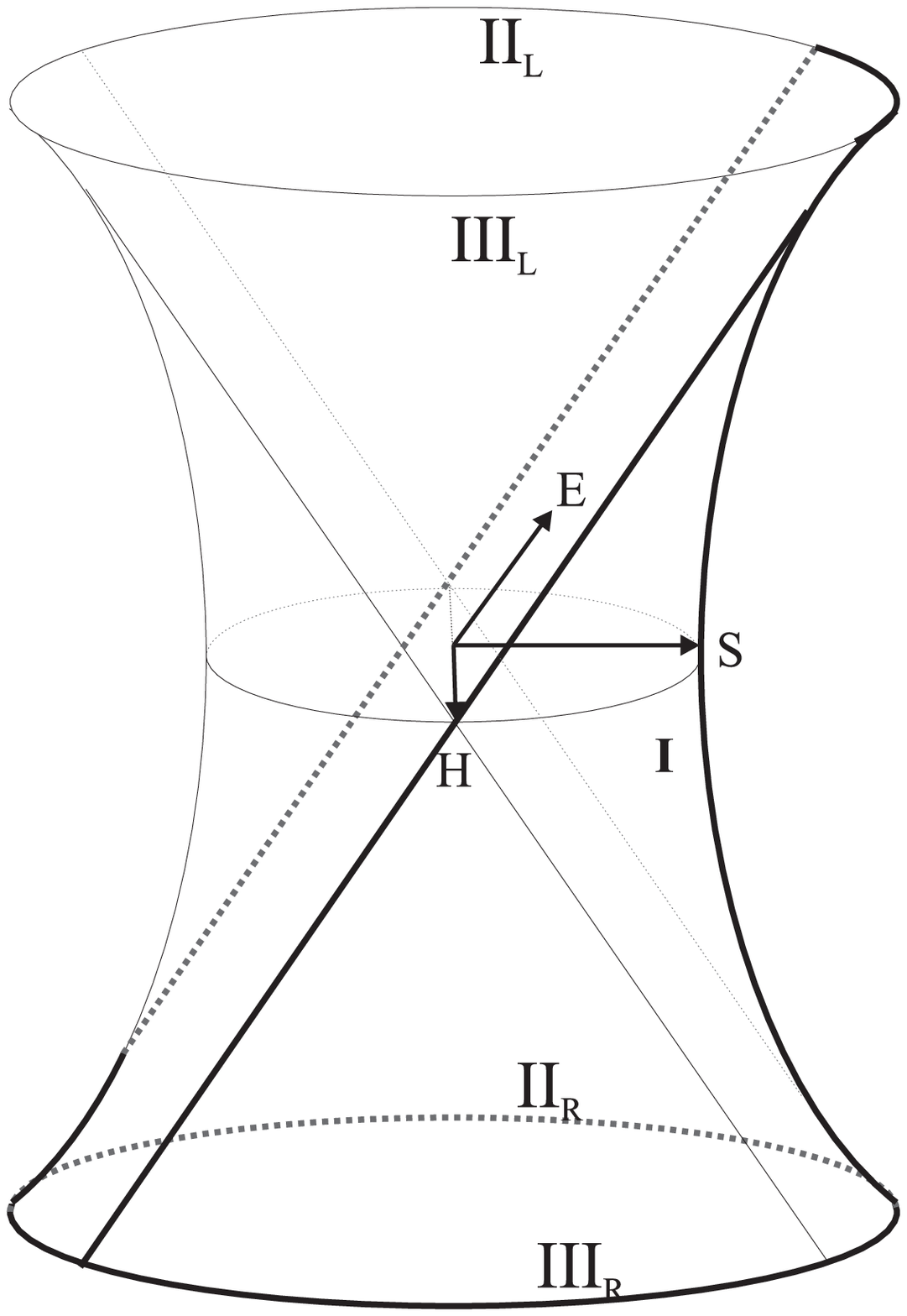}}
\caption[]{\label{OrbiteExt} The extension we consider, consisting
in each leaf in half a hyperboloid. This region exactly coincides
with the orbits of an $AN$ subgroup of $\SL$.}

\end{center}
\end{figure}

One can also look at the section by the $\theta = 0$ surface. One
then gets an usual two-dimensional Penrose diagram, where the
singularities and horizons are drawn, as displayed in
Fig.~\ref{Penrose2D}.














\subsubsection{Extended BTZ black holes}\label{Extensions}
\begin{figure}[hbt]
\begin{center}

{\epsfig{scale=0.7,file=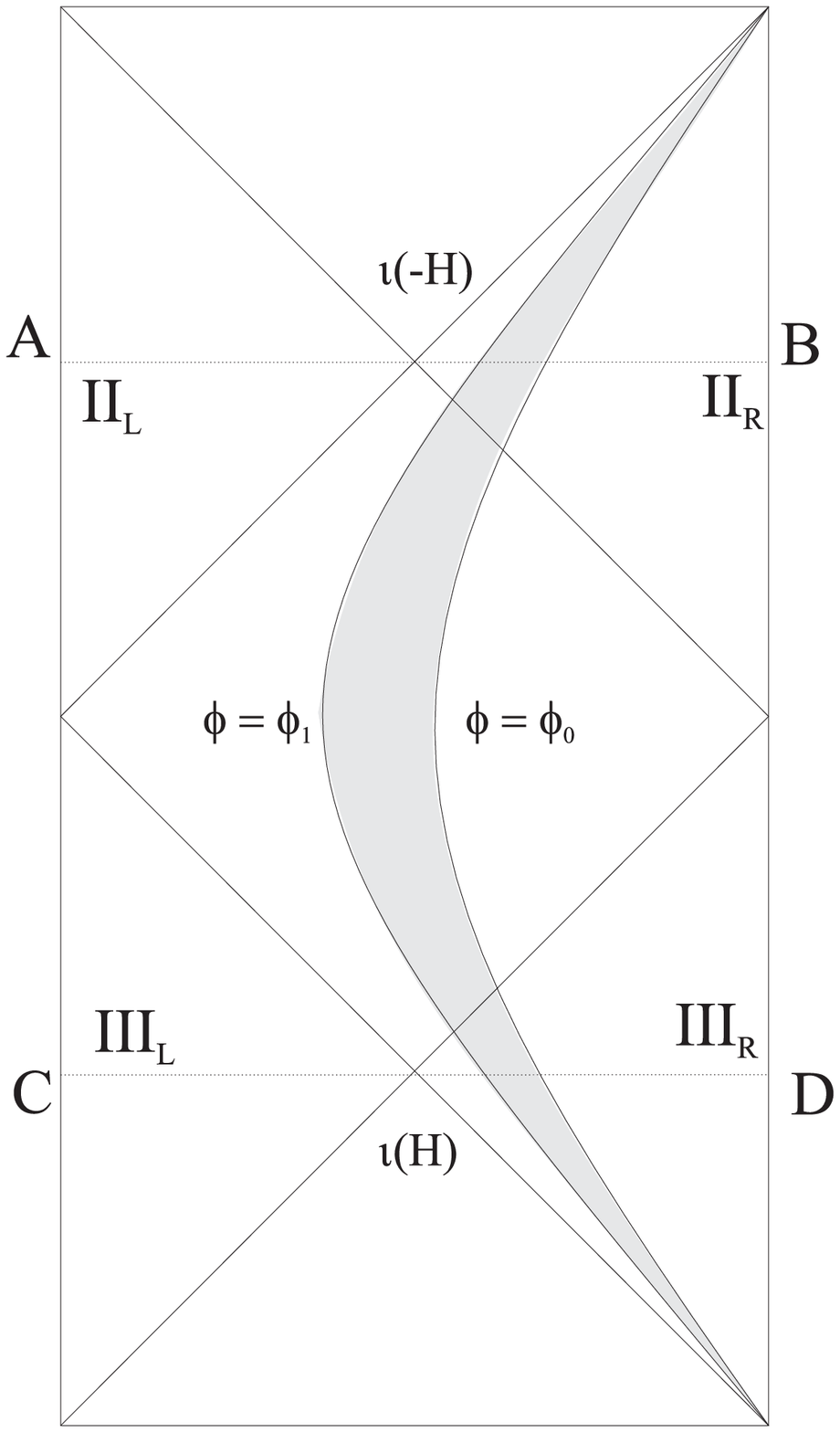}}
\caption[]{\label{feuillerhoExt} Same section as
Fig.\ref{feuillerho}, for the extension we considered. The shaded
region corresponds to a section of a fundamental domain of the
BHTZ action.}

{\epsfig{scale=0.7,file=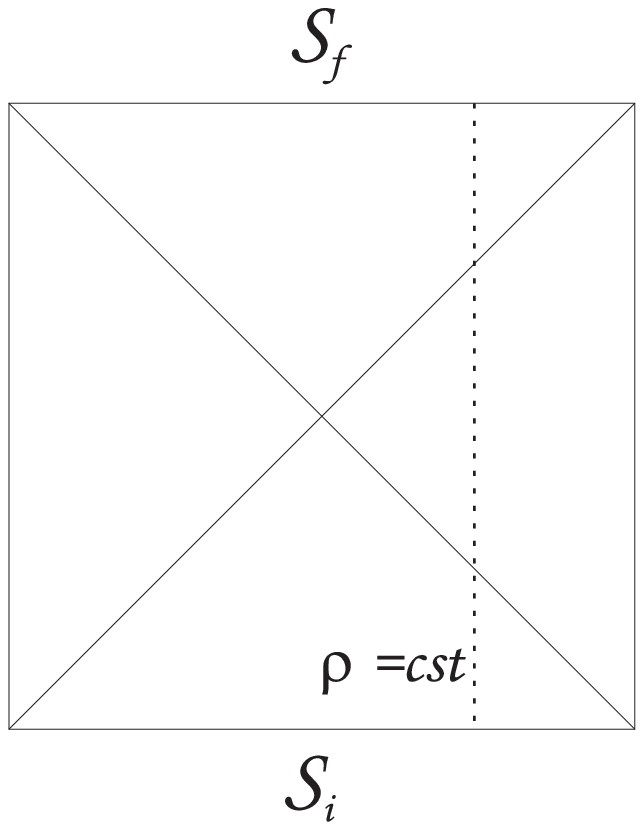}}
\caption[]{\label{D1brane} A $\rho=cst.$ D1-brane is represented
in this Penrose diagram of the non-rotating massive BTZ black
hole, where each point represents a circle. We see the brane
emerging from the past singularity, cross the horizons and end on
the future singularity. Note that a similar behavior is observed
for D0-branes in the $\SL/U(1)$ two-dimensional black hole (see
\cite{Yogendran}) }

\end{center}
\end{figure}

If we relax the condition that closed time-like curves must be
avoided, while keeping the condition that the quotient space $\SL
/ \psi_\Z$ be a Hausdorff manifold, there are two inequivalent
possibilities to extend the BTZ space-time. Either we choose to
consider, in each leaf, the region $I \cup II_L \cup III_R$,
either we take $I \cup II_R \cup III_R$ (see Figs.~\ref{OrbiteBD},
\ref{feuillerho},\ref{feuillerhoExt}). The latter extension
exhibits an interesting property : in this situation, each leaf in
the extension admits an action of the two-parameter $AN$ subgroup
of $\SL$. Indeed, we note that the region $I \cup II_R \cup III_R$
in $Ad(G)\h$ (see Fig.~\ref{OrbiteExt}) can be parameterized
as
\be \label{hyperbExt} X= Ad\left( \exp(\frac{\phi}{2}\h)\exp(w\e)
\right) \s \quad ,\quad -\infty<\phi<+\infty,\, -\infty<w<+\infty
\,, \ee

where $\s = Ad\left(\exp (-\frac{\pi}{4})\t)\right) \h$.

Therefore, if $x = \tau_{r \exp (-\frac{\pi}{4}\t)}\left(\exp(\rho
 \h)\right) \in I \,\cup\, II_R\, \cup\, III_R \,\subset {\cal O}_\rho$, where
 $r=\exp(\frac{\phi}{2}\h)\exp(w\e)$, then
 $\tau_{r'} (x) \in I \,\cup\, II_R\, \cup \,III_R\, \subset {\cal O}_\rho, \, \forall r'\in
 R=AN$.

A section by the $\rho=0$ surface is depicted in
Fig.~\ref{feuillerhoExt}, where two coordinate lines $\theta
=cst.$ are drawn. The region in between corresponds to a
fundamental domain of the BHTZ action. Hereafter, the maximally
extended BTZ space-time whose each leaf is constituted by the
regions $I \,\cup\, II_R\, \cup \,III_R$ will be denoted $EBTZ$.

\section{Non-commutative geometry from string theory}


\subsection{Open strings in flat space-time in presence of constant B-field}
There are actually several ways to observe the emergence of
non-commutativity by analyzing open string theory in flat
space-time in the presence of a constant magnetic field $B$ on a
D-brane. We will focus on one of them, namely the direct canonical
quantization. Consider an open string propagating in a flat
D-dimensional space-time ${\cal M}^d$ with metric $\eta_{\mu \nu}$
in presence of a constant $B_{\mu \nu}$. The string worldsheet,
i.e. the surface swept out by the string during its temporal
evolution, will be parameterized by a set of fields $X^\mu :
\Sigma \rightarrow {\cal M}^d$, where $\Sigma$ is the strip
$[0,\pi] \times \R$. The action of the bosonic open string theory
in this background geometry, in the conformal gauge, is
\begin{equation}\label{S}
 S = -\frac{1}{4 \pi \alpha'} \int_\Sigma \, d^2 x \, (\eta_{\mu\nu}
 + B_{\mu \nu}) \, \p_+ X^\mu \p_- X^\nu \quad,
\end{equation}
where the string tension

 is $T=\frac{1}{2
\pi \alpha'}$, $x_+$ and $x_-$ are light cone coordinates on the
worldsheet\footnote{ The string worldsheet can be parameterized in
various ways. Usually, one starts with coordinates $(\sigma,\tau)$
for the strip $[0,\pi] \times \R$, where $\p \Sigma$ corresponds
to $\sigma=0,\pi$. The light-cone coordinates are defined by
$x^\pm = \tau \pm \sigma$. It is sometimes useful to go from a
lorentzian worldsheet to an euclidean one by setting $t= i \tau$.
Another parametrization is $z=e^{i x^-}$, $\bar{z}=e^{i x^+}$, in
which the worldsheet is mapped to the upper half plane. For
euclidean worldsheet, $z$ and $\bar{z}$ are complex conjugated
variables, and $\p \Sigma$ is given by $z=\bar{z} = \pm e^t$, i.e.
the real line.}. The equations of motion for the string
coordinates following from the variation of action (\ref{S}) are
given by
\begin{equation}\label{EOM}
 \p_+\p_- X^\mu = 0 \quad , \quad \mu=0,\cdots,d-1 \quad
 \end{equation}
and are not affected by the presence of the constant B-field. By
introducing the currents
\begin{equation}\label{courantspm}
 J_+^\mu = \p_+ X^\mu \quad \mbox{and} \quad  J_-^\mu = \p_- X^\mu
 \quad,
\end{equation}
the equations of motion simply state that
\begin{equation}
 J_+ = J_+ (x^+) \quad \mbox{and} \quad  J_- = J_- (x^-) \quad,
\end{equation}
or, expressed in terms of  $z=e^{i x_-}$ and  $\bar{z}=e^{i x_+}$
(see footnote), that the currents
\begin{equation}\label{courantszz}
 J^\mu = \p X^\mu \quad \mbox{and} \quad  \bar{J}^\mu = \pb X^\mu
 \quad,
\end{equation}
are purely holomorphic or anti-holomorphic, respectively.
Variation of the action (\ref{S}) shows that the equations of
motion (\ref{EOM}) have to be supplemented by boundary conditions
at the boundary of the string worldsheet, which can be satisfied
in two different ways :
\begin{equation} \label{Dir}
 \p_\tau X^\mu_{|\sigma=0,\pi} =
 0 \quad .
 \end{equation}
 or
\begin{equation} \label{GenNeum}
 \left(\eta_{\mu \nu} \p_\sigma X^nu - B_{\mu \nu} \p_\tau X^\nu\right)_{|\sigma=0,\pi} =
 0 \quad,
 \end{equation}
The first one corresponds to Dirichlet boundary conditions, while
the second one corresponds to generalized Neumann conditions
(Neumann boundary conditions are recovered for $B=0$). We can
impose independently either of the conditions in each of the $d$
dimensions. Suppose we impose Dirichlet conditions for $\mu =
p,\cdots,d-1$ and Neumann conditions for $\mu = 0,\cdots,p-1$.
Then, the string coordinates satisfy
\begin{equation}
 X^\mu = X^\mu_0 \quad, \quad \mu = p,\cdots,d-1 \quad,
 \end{equation}
that is, the endpoints of the string are restricted to move on a
$p$-dimensional hyper-plane in ${\cal M}^d$. This hyper-plane is
called a {\itshape Dp-brane}. Let us now focus on the solutions to
(\ref{EOM}) along the brane. With boundary conditions
(\ref{GenNeum}), one finds (by setting $\alpha' = 2$ for
simplicity, see \cite{BranesCurved,DiVecchia,Johnson,Chu} for
details)
\begin{equation}\label{ModeExp}
X^\mu (z,\bar{z}) = x^\mu - i \alpha_0^\mu \ln z \bar{z} - i
B^\mu_\nu \, \alpha_0^\nu \ln \frac{z}{\bar{z}} + i
\underset{n\ne0}{\sum} \frac{\alpha^\mu_n}{n} (z^n + \bar{z}^{-n})
+  i \underset{n\ne0}{\sum} \frac{B^\mu_\nu \alpha^\nu_n}{n} (z^n
+\bar{z}^{-n})
\end{equation}
The theory can be quantized by imposing equal time canonical
commutation relations between the $X^\mu$'s and their conjugate
momenta (see e.g.\cite{Johnson} p38). This amounts to take the
boundary condition as constraints and perform the canonical
quantization.  Then, $x^\mu$ and $\alpha_n^\mu$ become operators
satisfying
\begin{equation}
 [\alpha_n^\mu , \alpha_m^\nu] = n G^{\mu\nu}\delta_{n,-m} \quad
 ,\quad
 [x^\mu , \alpha_n^\nu] = i \sqrt{\alpha'} G^{\mu\nu}\delta_{0,n}
\end{equation}
\begin{equation}\label{xmunu}
   [x^\mu , x^\nu ] = i \Theta^{\mu\nu} \quad ,
  \end{equation}
where $G^{\mu \nu} = \left( \frac{1}{\eta + B} \, \eta \,
\frac{1}{\eta - B}\right)^{\mu\nu}$ and $\Theta^{\mu\nu}= \left(
\frac{1}{\eta + B} \, B \, \frac{1}{\eta - B}\right)^{\mu\nu}$. It
follows from (\ref{xmunu}) that at the boundary, we have
\begin{equation}
 [X^\mu (\tau), X^\nu (\tau) ] = i \Theta^{\mu \nu} \quad ,
 \end{equation}
 which implies that the D-brane worldvolume, where the open string
's endpoints live, is a noncommutative manifold.
The same result can be obtained from the open string Green
function (see \cite{SeibWitt}) or by analyzing the modification of
the operator product expansion of the open string vertex operators
due to the boundary B-term \cite{DefQ}.
\subsection{Strings on group manifolds : the WZW model}
One of the features making strings attractive is that they
exhibit, upon quantization, a spin-2 particle identified with the
graviton in their spectrum. A common approach to tackle String
Theory is called {\itshape perturbative}, in the sense that it
consists in studying the propagation of strings in non-trivial,
fixed, curved backgrounds. From a conceptual point of view, this
is quite analogous to what can be done in electromagnetism, by
using the photon description to describe processes in strong,
non-perturbative magnetic fields, as the ones around a pair of
Helmholtz coils \cite{Johnson}. The background is then viewed as a
"coherent state" of elementary particle quanta. However, the story
is a little bit more complicated than just replacing everywhere,
in the preceding section, $\eta_{\mu\nu}$ by any curved metric
$g_{\mu\nu}$ along with any B-field. The background fields are
indeed expected to satisfy conditions in order to represent an
"exact string background". These conditions are expressed by the
following. The action (\ref{S}) may alternatively be viewed as a
two-dimensional field theory, with $d$ fields $X^\mu
(\tau,\sigma)$, the metric and B-field being reinterpreted as
(field-dependent) coupling constants. The metric $h_{\alpha\beta}$
on the two-dimensional manifold (worldsheet) has no local
propagating degrees of freedom, in the sense that it can always be
brought to the flat two-dimensional metric $\eta_{\alpha \beta}$,
due to the reparametrizations\footnote{i.e. transformations
$x^\alpha \rightarrow f^\alpha(x)$}  and Weyl\footnote{i.e. local
rescalings of the metric tensor $h_{\alpha\beta} \rightarrow
\Lambda(x)h_{\alpha\beta}$} invariances of the action (\ref{S})
The conformal gauge choice corresponds to bringing
$h_{\alpha\beta}$ to $e^\phi h_{\alpha\beta}$ for some function
$\phi$. In this gauge, the string action is still invariant under
some residual local symmetries, corresponding to infinitesimal
conformal transformations. This leads us to the observation that
string theory in the conformal gauge is equivalent to a
two-dimensional conformal field theory. In curved backgrounds,
this requirement survives, and the preservation of the Weyl
invariance imposes equations to be satisfied by the background
field (the beta-functionals for the couplings in the 2-D field
theory have to vanish; more details can be found in \cite{GSW},
\cite{Johnson}p56, \cite{Ooguri}pp17-25, \cite{Pol}).

Besides the flat case, an important class of exact string
backgrounds is provided by the Lie group manifolds. Let $G$ be a
Lie group, endowed with an invariant metric (usually taken as the
Killing metric) on its Lie algebra ${\cal G}$, denoted by $<-,->$.
Let $g : \Sigma \rightarrow G$ be a map expressing the embedding
of the string worldsheet into the group manifold. The action for
an open string ending on a D-brane $Br$ in the group manifold is
given by the {\itshape boundary Wess-Zumino-Witten model} (BWZW),
whose action reads
\begin{equation}\label{BWZW}
S^{BWZW} = \int_\Sigma \, \langle g^{-1} \p_ig,g^{-1} \p^i g
\rangle + \int_{{\cal M}} H - \int_D \omega \quad.
\end{equation}
${\cal M}$ is a three-dimensional submanifold of $G$, such that
$\p {\cal M} = g(\Sigma) \cup D$. Because for an open string
$\p\Sigma \ne \emptyset$, $D$ is to be chosen as a two-dimensional
submanifold included in the brane $Br$ (where the open string
ends) such that $\p (g(\Sigma) \cup D)=0$.  $H = \langle \theta,
[\theta,\theta] \rangle$ denotes a closed three-form field
strength on the target group manifold, where $\theta = g^{-1}dg$
stands for the left-invariant Maurer-Cartan one-form on $G$, while
the two-form on $Br$ is defined by $d\omega = H_{|Br} :=
dB_{|Br}$. By choosing coordinates on the group manifold, this
action can be rewritten as (\ref{S}), with $g_{\mu\nu} =
\beta_{ab}\theta^a_\mu\theta^b_\nu$, $\beta_{ab}$ denoting the
components of the Killing metric on $G$, and a certain
(non-constant) B-field\footnote{Actually, (\ref{BWZW}) contains an
additional coupling at $\p \Sigma$, represented by a term
$\int_{\p \Sigma} \, A_\mu \p_\tau X^\mu d\tau$, which can be
recast as $\int_{\Sigma} \, F_{\mu\nu} \p_+ X^\mu \p_- X^\nu
d^2x$, with $F=dA$, and then included in the definition of B}. We
do not discuss here the ambiguities resulting in the choices of
${\cal M}$ and $D$ in the two topological terms of (\ref{BWZW})(
for more details, see \cite{sttalk,stD0,st3,Rib} and references
therein). By considering only the first two terms in the action,
one gets the usual WZW model \cite{Gep-Witt,NonAbel} describing
closed strings on group manifolds. The geometric meaning of the
field $H$ can be understood as a parallelizing torsion, added to
the metric connection to make it flat. It is in fact the existence
of this torsion in the case of group manifolds that constitute a
crucial ingredient in making the two-dimensional nonlinear sigma
model conformally invariant \cite{BCZ85Rib}.

 Let us now have a
look at (\ref{BWZW}). This model shares actually many features
with the flat situation we considered in preceding section. The
variation of this action w.r.t. the field $g$ yields the equations
of motion
\begin{equation}
 \p \bar{J} = 0 \quad , \quad  \pb J = 0 \quad,
 \end{equation}
 where
 \begin{equation}
  J =g^{-1}\p g  \quad \mbox{and} \quad \bar{J} = -\pb g g^{-1}
  \quad .
  \end{equation}
  Thus, as in the flat case (see (\ref{courantszz})), the theory possesses purely
  holomorphic and anti-holomorphic currents. The WZW-model is
 invariant under the {\itshape current algebra} which
  stems from the invariance of the action under
  \begin{equation}
   g(z,\bar{z}) \rightarrow \Omega(z) g(z,\bar{z})
   \bar{\Omega}(\bar{z})\quad .
   \end{equation}
   This infinite-dimensional symmetry is precisely generated by
   the conserved currents, whose modes generate two commuting
   affine Kac-Moody algebras. Conformal
   invariance of the model can actually be deduced from its
   invariance under the current algebra, through the Sugawara
   construction \cite{KNZ,Gep-Witt,DiFr}.

   By analogy with (\ref{Dir}) and (\ref{GenNeum}), which can further be rewritten
   as
\begin{equation}
 J = - \bar{J} \quad (D) \quad, \quad J = C \bar{J} \quad (N) \quad
 \mbox{at}\quad
 z=\bar{z} \, ,
 \end{equation}
 with $C^\alpha_\beta = \left((\eta - B)^{-1}\right)^\alpha_\mu
 \left( \eta + B \right)^\mu_\beta$,
    a particular
   class of D-branes in WZW models is obtained by imposing
   {\itshape gluing conditions} on the chiral currents :
   \begin{equation}\label{gluing}
    J = R \bar{J} \quad,
    \end{equation}
    where $R$ is a metric preserving Lie algebra automorphism.
These D-branes are called {\itshape symmetric}, because they
describe configurations preserving the maximal amount of symmetry
of the bulk theory, that is, conformal invariance and half of the
current algebra \cite{Stanciu}. It is important to note that these
conditions {\itshape do not} correspond, a priori, to boundary
conditions. It is only {\itshape a posteriori} that the last term
of (\ref{BWZW}), which does not affect the bulk equations of
motion, is added and must be fitted to ensure that the boundary
conditions extracted by varying the action coincide with the ones
coming from the gluing conditions.

The gluing conditions can now be used to determine the geometry of
the resulting D-branes\footnote{which can be seen, in the
semi-classical limit, as submanifolds of $G$} \cite{Stanciu,stD0}.
 Because the string's worldsheet ends on the D-brane, the vectors
 $\p_\tau g$ at $\p \Sigma$ must be tangent to the D-brane. From
 (\ref{gluing}), we get
 \begin{equation}
  (\mathbb{I} + R \circ Ad(g))g^{-1} \p_\tau g  =
  (\mathbb{I} - R \circ Ad(g))g^{-1} \p_\sigma g  \quad \mbox{at}
  \quad \p \Sigma .
  \end{equation}
  This implies that the vector $g^{-1} \p_\tau g$ lies in the
  image of the operator $(\mathbb{I} - R \circ Ad(g))$. Thus,
  \begin{equation}
   \p_\tau g =Y g  - g R(Y) \quad,
   \end{equation}
   for some $Y \in {\cal G}$.
   This means that $\p_\tau g$ must be tangent to a {\itshape
   twisted conjugacy class}. A twisted conjugacy class of an
   element $g\in G$ is defined by(see also (\ref{orbit}))
   \begin{equation}\label{cctw}
    C_R (g) = \{ h g r(h^{-1}) \quad | \quad g\in G\} \quad,
  \end{equation}
where the map $r:G \rightarrow G$ is defined by
\begin{equation}
 r(e^{tX}) = e^{tR(X)} \quad,
 \end{equation}
for $t$ small enough and $X\in{\cal G}$. If $G$ is connected, then
$r$ extends to a Lie group automorphism. In conclusion, D-branes
corresponding to (\ref{gluing}) can be identified to regular
($R=Id$), translated ($R =$ inner automorphism) or twisted ($R =$
outer automorphism) conjugacy classes in $G$. Finally, let us
notice that the gluing conditions can indeed by interpreted as
boundary conditions stemming from the variation of (\ref{BWZW}),
with the two-form field on a twisted conjugacy class uniquely
determined by the automorphism $R$ \cite{AlexSchom,Stanciu,stD0}
 :
\begin{equation}
 \omega_g (u,v) =  \langle g^{-1}u , \frac{\mathbb{I} + R \circ
 Ad(g)}{\mathbb{I} - R \circ Ad(g)} g^{-1}v \rangle \quad ,
 \end{equation}
 where $u,v \in T_g^{C_R(g_0)}$.

We now particularize the discussion to the case of interest here,
namely $G=\SL$ (see (\ref{AdSl})).   Symmetric D-branes of the
$Sl(2,\R)$ WZW model are of three types : two-dimensional
hyperbolic planes ($H_2$), de Sitter branes ($dS_2$) and anti-de
Sitter branes ($AdS_2$).  It was shown in \cite{BachPetr}, that
the $AdS_2$ worldvolumes, corresponding to twisted conjugacy
classes, are the only physically relevant classical
configurations.

We now make a link with the geometry of the non-rotating BTZ
black-hole. We saw in the first section that the spinless BTZ
black hole admits a foliation by leaves, the $\rho$ = constant
surfaces, which are stable under the action of the BHTZ subgroup
and constitutes twisted conjugacy classes in $Sl(2,\R)$ ($AdS_2$
spaces). From our previous discussion, these solutions correspond
to projections of twisted $AdS_3$ conjugacy classes that wrap
around BTZ space, and they may be interpreted as closed DBI
1-branes in BTZ space. These branes are represented in
Fig.~\ref{D1brane}.

There is still one point that would be needed to be discussed,
namely if there is an analog, in the context of WZW models, of the
noncommutative structures that emerge in the flat case (actually
the Moyal product). The answer is not quite clear at present. A
straightforward interpretation in terms of coordinates on the
group manifold seems no longer possible since we do not have an
analog of (\ref{ModeExp}), and since the gluing conditions are
expressed in terms of the currents, while these have not a simple
relation with the coordinates as in the flat case ($J^\mu =\p
X^\mu + \cdots$, where $\cdots$ stand for nonlinear terms in the
coordinates). In the case of $G=SU(2)\simeq S^3$, it was observed
in \cite{AleksReckSch} that D-branes could be described by
nonassociative deformations of fuzzy spheres, by analyzing the
vertex operator algebra. Recently, D-branes in Nappi-Witten
backgrounds, corresponding to WZW models based on the Nappi-Witten
groups\footnote{In 4 dimensions, this group is a central extension
of the two-dimensional euclidean group $ISO(2) = SO(2) \times
\R^2$}, were also shown to exhibit non-commutativity \cite{Sam}.

In the case of $AdS_3$ and the BTZ black hole (corresponding to a
WZW model based on a semi-simple {\itshape and} non-compact
group), the question has not been investigated yet. In the next
section, we discuss deformations of D-branes in extended BTZ
space-times, which could reveal relevant for open string theory on
BTZ black holes space-times.

\section{Strict deformation of extended BTZ space-times}
\subsection{The purpose of deformation}
Historically, star product theory was introduced as an autonomous
and alternative formulation of Quantum Mechanics. In this
framework, the quantization of a classical system whose
configuration space is $N$ is expressed as a correspondence (the
{\itshape quantization map}) between {\itshape classical
observables} (functions defined on the phase space $T^*N$)
 and {\itshape quantum observables} (operators acting on a Hilbert
 space, the {\itshape state space}). A star product allows to read
 the composition of the quantum operators at the level of
 functions on the phase, without any reference to a Hilbert space
 representation.

As an example, consider Weyl's method for quantizing a free
particle on a line. The phase space of this system is $\R^2 =
\{(q,p)\}$. The quantization map is the linear map

\be W_{\lambdabar} : {\cal S} (\R^2) \rightarrow \mathcal{B} (L^2
(\R)) \quad, \ee
 relating Schwartz's functions
  $f: \R^2 \rightarrow \R$ on the phase
space to linear bounded operators on Hilbert space of square
integrable functions on the configuration space, $L^2 (\R)$. The
{\itshape Weyl transform} is given by
\begin{equation}\label{Weyl}
[(W_{\lambdabar} u)f] (q) = \frac{1}{\lambdabar} \int_{\R \times
\R} e^{\frac{i}{\lambdabar} p (q-q')} u(\frac{q+q'}{2},p)f(q') \,
dq' dp
\end{equation}

for $f \in L^2 (\R), u \in C^{\infty}(\R^2)$ and $q,q',p \in \R$.
The Moyal-Weyl product of two functions $u,v \in C^{\infty}(\R^2)$
is formally defined as
\begin{equation} \label{starOp}
W_{\lambdabar} (u \star_{\lambdabar} v) = W_{\lambdabar}(u) \circ
W_{\lambdabar}(v)
\end{equation}
The Moyal-Weyl product can be written in integral form as
\begin{equation}\label{IMP}
(u \star_{\lambdabar} v)(x) = \frac{1}{2 \pi\lambdabar} \int_{\R^2
\times \R^2} u(y)v(z) e^{- \frac{i}{\lambdabar} (x-z) \wedge
(y-x)} d^2 y d^2 z
\end{equation}
for $x,y,z \in \R^2, u,v \in C^{\infty}(\R^2)$ and where $x \wedge
y=\omega^0 (x,y), \omega^0$ being the symplectic 2-form on the
phase space $\R^2$\footnote{$\omega((q_1,p_1),(q_2,p_2)) = q_1 p_2
- q_2 p_1$}. Interpreting formula (\ref{IMP}) as an oscillatory
integral with parameter $-\frac{1}{\lambdabar}$, one can use a
stationary phase method
 to obtain the following asymptotic expansion :
\begin{equation}\label{DMP}
(u \star_{\lambdabar} v)(x) = u(q,p) \exp[i
\lambdabar(\overleftarrow{\partial_q}\overrightarrow{\partial_p} -
\overleftarrow{\partial_p} \overrightarrow{\partial_q})] v(q,p)
\end{equation}
To first order in the parameter $\lambdabar$, we get \be u
\star_{\lambdabar} v \sim u.v + i\, \lambdabar \{u,v\} +
O(\lambdabar^2) \quad, \ee

where $\{\,,\,\}$ is the Poisson tensor associated to $\omega^0$.
The Moyal-Weyl product appears as a one-parameter deformation of
the usual commutative pointwise multiplication of functions in the
direction of the classical Poisson bracket.

More generally, consider the Poisson algebra
$\left(C^\infty(M),\{\,,\,\}\right)$ of smooth functions on a
manifold $M$ endowed with the usual pointwise product. A star
product on $M$ is a $\R[[\lambdabar]]$-bilinear map

\be C^\infty(M)[[\lambdabar]] \times C^\infty(M)[[\lambdabar]]
\rightarrow C^\infty(M)[[\lambdabar]] : (f,g) \rightarrow f \star
g \quad, \ee where $f$ and $g$ are formal power series in
$\lambdabar$ with functions $C^\infty(M)$ as coefficients. The
following conditions are to be satisfied :
\begin{itemize}
{\item $f \star_{\lambdabar} g = f.g + i \lambdabar \{f,g\} +
O(\lambdabar^2)$}

 {\item (f $\star_{\lambdabar}g) \star_{\lambdabar} h =
  f\star_{\lambdabar}(g\star_{\lambdabar}h) \quad \forall f,g,h\in C^\infty(M)$}

{\item $1 \star_{\lambdabar} f = f \star_{\lambdabar} 1 =f \quad
\forall f\in C^\infty(M)$}
\end{itemize}

The Moyal-Weyl product enjoys the property to be {\itshape
strict}, meaning that in a suitable functional framework, the
product of two functions is again a function, rather than a
{\itshape formal} power series in the deformation parameter, as in
star product theory (in the strict sense). We will mainly be
concerned in the two next sections in constructing strict
deformations of the algebra of functions in the extended BTZ
space-time. The fact that every leaf of the foliation admits an
action by a two-parameter solvable Lie group will reveal the
crucial ingredient to achieve this.

\subsection{Universal deformation formula for group
actions}\label{GroupAction} In this section, we will show how to
construct a star product on a manifold admitting an action of a
Lie group $G$, from an invariant star product on $G$. The
procedure can be seen as a generalization of Rieffel strict
deformation theory for actions of $\mathbb{R}^d$ \cite{Rieffel}.
Here we follow \cite{PierreStrict,Pierre2} and section 4 of
\cite{StarP}.

Let $G$ be a Lie group. We define the left (resp. right) action,
$L^\star_{g}$ (resp. $R^\star_{g} $), on ${\rm Fun}[G,\CC]$ as

\be L^\star_{h}[f](g) = f(hg), \quad R^\star_{h}[f](g) = f(gh)
\quad, \forall g,h\in G\quad. \ee

Suppose there exist a left action $\tau$ of $G$ on a manifold M,
i.e. :

\be \tau : G \times M \rightarrow M : (g,x)\rightarrow \tau_g(x)
\quad, \, \tau_g \circ \tau_h = \tau_{gh}. \ee

This action on $M$ induces an action $\alpha$ on ${\rm
Fun}[M,\CC]$ :
 \beq
 \alpha : G \times {\rm
Fun}[M,\CC] \rightarrow {\rm Fun}[M,\CC] : (g,u)\rightarrow
\alpha_g[u], \quad \nonumber\\
\mbox{with}\quad  \alpha_g[u](x) = u(\tau_{g^-1}(x)), \quad
\alpha_g \circ \alpha_h = \alpha_{gh}, \, \forall x \in M. \eeq

For fixed $x\in M$, we may define a map
         $\tilde\alpha^{x}$ from ${\rm Fun}[M,\,\CC]$ into ${\rm
         Fun}[G,\,\CC]$:
\begin{eqnarray} {\rm Fun}[M,\,\CC]\rightarrow
\kern-1.2em^{\tilde\alpha^{x}}\kern0.5em{\rm Fun}[G,\,\CC]:
         \ u \mapsto \tilde\alpha^{x}[u]&&\\
         \forall\  g \in G\ :\ \tilde\alpha^{x}[u](g)=u(\tau_{g^{-1}}(x)) &&.
         \end{eqnarray} Let us also assume that on $G$ we have a left invariant
         star product, denoted by~$\st LG$, {\it i.e.} a star product
         satisfying the relation
\begin{equation} L^\star_{g} [f_1 \st LG f_2]=L^\star_{g} [f_1]\st LG
         L^\star_{g} [f_2] \qquad .
         \end{equation}
         From the left invariant star product on $G$, we induce a star
         product on $M$, denoted $\st {\ \ }M$, defined as :

         \begin{equation}\label{babel}
         \left(u \st {\ \ }M v\right)(x):=\left(\tilde\alpha^{x}[u]\st LG
         \tilde\alpha^{x}[v]\right)(e)\qquad,
         \end{equation}
         $e$ denoting the identity element of $G$. One can check
         that all the required properties are indeed satisfied,
         essentially because of the properties of $\st LG$.

\subsection{Extended BTZ and $AN$-invariant star-products}
We know from section \ref{Extensions}, that each leaf, in $EBTZ$,
admits an action of $AN \subset \SL$. Therefore, from section
\ref{GroupAction}, it remains to us to find a left-invariant star
product on this group. This was first carried out in
\cite{PierreStrict}. Another, more heuristic, approach was
considered in \cite{StarP}. We start from the general expression

\begin{equation} \label{SP}
         (u*v)(x)= \int {\rm K}[x,y,z]\,u(
         y)\,v( z)\,d\mu_y\,d\mu_ z,
         \end{equation} where the (left invariant) measure used is simply
         $d\mu_x= da_ x\ dn_x$ and where $x_i = (a_{x_i},n_{x_i})
         \in AN \cong \R^2$.

Writing

\be K[x,y,z] = B(x,y,z) \, \exp (\frac{i}{\lambdabar} \Psi(x,y,z))
\quad , \ee

and imposing the left-invariance, the associativity, the existence
of left and right units and the right asymptotic expansion (the
Poisson bracket on $AN$ being defined from the canonical
left-invariant symplectic structure) constrains the form of the
amplitude and phase, which can be explicitly obtained (see
\cite{StarP} for details). One finds for the phase function
\begin{equation}
         \Psi= \left\{ \sinh\left[(a_y-a_x)\right]\, n_z + \sinh\left[(a_z-a_y)\right]\, n_x
         + \sinh\left[(a_x-a_z)\right]\, n_y  \right\}\label{phase}\qquad
         ,
         \end{equation}
while a class of amplitudes corresponds to
  \begin{equation}
         B=\frac{1}{(2\,\pi\,\lambdabar)^2 }
         \frac{{\cal P}(a_y-a_x){\cal P}(a_x-a_z)}{{\cal P}(a_z-a_y)}\cosh(a_z-a_y)\qquad
         ,
         \end{equation}
where ${\cal P}$ is any real even function satisfying ${\cal P}(0)
= 1$.

\section{Summary and outlook}
In this talk, we put forward some geometrical properties of
non-rotating massive BTZ black holes, relevant both from the
string theory point of view as well as in the context of
non-commutative geometry (strict deformations). We showed that
$AdS_3$ space, from which the black hole is constructed by
discrete identifications, admits a foliation by two-dimensional
twisted conjugacy classes, all diffeomorphic, and stable under the
identifications. We then argued that these surfaces constitute
{\itshape winding D1-branes} in the BTZ black hole space-time,
seen as an exact string background via the WZW model. By extending
the space-time across the chronological singularities, we observed
that each leaf of the foliation admitted an action of a minimal
parabolic subgroup of $\SL$, the solvable Lie group $AN$. This
allowed us to construct a strict deformation of the algebra of
functions on these spaces, following a Rieffel-like construction.

At this point, the relation between our construction and string
theory, in the spirit of the emergence of the Moyal product from
open string theory in flat space, is far from clear. This comes
essentially from the fact that string theory on BTZ black holes,
though in principle tractable using the WZW model, is poorly
understood. It is worth noticing that it is only recently
\cite{MO} that the $\SL$ WZW model, describing strings on $AdS_3$,
has revealed to be consistent, from its physical content (an
infinite tower of massive states), as well as from the unitarity
of the quantum theory (no-ghost theorem) and the modular
invariance of its partition function. It is worth noticing that
the origin of this problem can be traced back to the early 90's
\cite{Moham,Petro,Hwang}.... The spectrum of string theory on BTZ
black hole can, in principle, be obtained from the spectrum on
$AdS_3$, by keeping only those states invariant under the
identifications, and by adding winding (spectral flowed) sectors.
However, it is not really known how to implement the
representations of the affine algebra that would correspond to
those winding sectors (see however \cite{TroostWin,HK}), nor how
to construct a modular invariant partition function. One obstacle
in achieving this is rooted in the need of working in a hyperbolic
basis of $\SL$, diagonalizing the action of the BHTZ subgroup,
thereby generating all the subsequent difficulties in dealing with
continuous bases. In particular, the affine characters of $\SL$ in
a hyperbolic basis, which could be relevant in writing down the
partition function, are not known (but see \cite{SLHypCh}).
Finally, in the case of the $SU(2)$ WZW model, where the spectrum
and a modular invariant partition function are known (see
\cite{DiFr}) the emergence of non-commutativity is rather trickier
than in the flat case \cite{AleksReckSch}.

 \section*{Acknowledgments}
        S.D. would like to thank the organizers of the "8\`emes
        Rencontres Math\'ematiques de Glanon" for giving him the opportunity
        to present this work,  as well as all the participants for the nice
        atmosphere and exchanges during this week. He is also
        grateful to M. Musso for his kind hospitality! Thanks to
        Michel Herquet for the permission of using Fig.3  and to Denis Haumont for the help in the
        conception of the other figures.

\newpage
\section*{Figures}

\newpage

\newpage

 \newpage

\newpage


\newpage
\begin{thebibliography}{999}
\bibitem{DT}  S. Deser, R. Jackiw and Gerard 't Hooft, Annals Phys.152:220,1984;
S. Deser and R. Jackiw, Annals Phys.153:405-416,1984

\bibitem{BTZ} M. Ba\~nados, C. Teitelboim, J. Zanelli,
         Phys. Rev. Lett. {\bf{69}} (1992) 1849, hep-th/9204099
         \bibitem{BHTZ} M. Ba\~nados, M. Henneaux, C. Teitelboim, J. Zanelli, Phys. Rev.
         {\bf{D48}} (1993) 1506, gr-qc/9302012

\bibitem{Wald}   R. M. Wald, {\itshape General Relativity}, University
of Chicago Press, 1984.


\bibitem{Barut} A. Barut, P. Raczka, {\itshape Theory of Group
Representation and applications}, World Scientific

\bibitem{Helgason}
S. Helgason, {\itshape Differential Geometry, Lie Groups, and
Symmetric Spaces},
 Academic
Press.(1978)


 \bibitem{BRS} P. Bieliavsky, M. Rooman, Ph. Spindel, Nucl.Phys. B645 (2002)
         349-364, hep-th/0206189

\bibitem{HE} S.W. Hawking, G.F.R. Ellis, {\itshape The large scale structure
         of space-time}, Cambridge Monographs on Mathematical Physics,
         Cambridge University Press (1973)

 \bibitem{BranesCurved} V. Schomerus, {\itshape Lectures on branes in curved
         backgrounds}, Class.Quant.Grav. 19 (2002) 5781-5847,
hep-th/0209241
 \bibitem{DiVecchia} Paolo Di Vecchia and Antonella Liccardo,
 {\itshape D branes in string theory,I}, NATO Adv.Study Inst.Ser.C.Math.Phys.Sci. 556 (2000) 1-59
, hep-th/9912161

\bibitem{Johnson} C. Johnson, {\itshape D-brane primer},
hep-th/0007170 ; {\itshape D-branes}, Cambridge Monographs on
Mathematical Physics


\bibitem{Chu}C.-S. Chu, hep-th/0502167

 \bibitem{SeibWitt} N. Seiberg, E. Witten, JHEP 9909:032, 1999,
         hep-th/9908142
\bibitem{DefQ} V. Schomerus, JHEP 9906 (1999) 030,
         hep-th/9903205
\bibitem{GSW} M.B. Green, J.H. Schwartz, E. Witten, {\itshape Superstring
         Theory}, Vol.1, Cambridge Monographs in Mathematical Physics,
         Cambridge University Press (1987)
\bibitem{Ooguri}Hirosi Ooguri, Zheng Yin
, {\itshape TASI Lectures on Perturbative String Theories},
hep-th/9612254

\bibitem{Pol} J.Polchinsky, {\itshape String theory}, Vols. 1 et
2 ,Cambridge Univ. Pr. (1998)

\bibitem{sttalk}Sonia Stanciu, Fortsch.Phys. 50 (2002) 980-985, hep-th/0112130

\bibitem{stD0}Sonia Stanciu,  JHEP 0010 (2000) 015, hep-th/0006145

\bibitem{AleksReckSch} Anton Yu. Alekseev, Andreas Recknagel, Volker Schomerus
, Mod.Phys.Lett. A16 (2001) 325-336 ,  hep-th/0104054; JHEP 0005
(2000) 010, hep-th/0003187; JHEP 9909 (1999) 023, hep-th/9908040

\bibitem{Sam} S. Halliday, R. Szabo, hep-th/0502054

\bibitem{st3}Sonia Stanciu, JHEP 9909 (1999) 028, hep-th/9901122

\bibitem{Rib}Sylvain Ribault, PhD thesis, hep-th/0309272

\bibitem{Gep-Witt} D.Gepner, E.Witten, Nucl.Phys.B278 (1986) 493-549
\bibitem{NonAbel} E. Witten, Commun.Math.Phys.92:455-472,1984
\bibitem{BCZ85Rib}  E. Braaten, T. Curtright, C.K. Zachos, Nucl.Phys.B260:630,1985

\bibitem{KNZ} V.G. Knizhnik, A.B. Zamolodchikov, Nuclear Physics
B247 (1984) 83-103
\bibitem{DiFr}P. Di Francesco, P. Mathieu, D. Senechal, {\itshape Conformal Field
Theory},  Springer (1997)

 \bibitem{Stanciu} S Stanciu, JHEP 0001 (2000) 025, hep-th/9909163
\bibitem{AlexSchom} A.Y. Alekseev, V. Schomerus, Phys.Rev. D60 (1999)
         061901, hep-th/9812193
\bibitem{BachPetr} C. Bachas, M. Petropoulos, JHEP 0102 (2001) 025,
         hep-th/0012234


\bibitem{Rieffel} M.A. Rieffel, Mem. Amer. Math. Soc., 106(506)(1993)

\bibitem{Pierre2} P. Bieliavsky, M. Bordemann, S. Gutt, S. Waldmann, Rev. Math
         Phys. {\bf{15}} (2003) 425, math.QA/0202126

\bibitem{PierreStrict} P. Bieliavsky, Journal of Symplectic Geometry Vol.1 N.2(2002)269, math.QA/0010004

\bibitem{StarP}P. Bieliavsky, S. Detournay, M. Rooman, Ph. Spindel, JHEP06(2004)031,
      hep-th/0403257
\bibitem{MO} J.Maldacena and H.Ooguri, J.Math.Phys, 42 (2001) 2929 , hep-th/0001053 ; J. Maldacena, H. Ooguri, J.
Son,  J.Math.Phys. 42 (2001) 2961-2977, hep-th/0005183 ;
J.Maldacena and H.Ooguri, Phys.Rev. D65 (2002) 106006,
hep-th/0111180

\bibitem{Moham}J. Balog, L. O'Raifeartaigh, P. Forgacs, A. Wipf,  Nucl.Phys.B325:225,1989
\bibitem{Petro}  P.M.S. Petropoulos, Phys.Lett.B236:151,1990
\bibitem{Hwang}  S. Hwang, Nucl.Phys.B354:100-112,1991

\bibitem{TroostWin} J. Troost, JHEP 0209 (2002) 041,
hep-th/0206118
\bibitem{HK} S.Hemming and E.Keski-Vakkuri, Nucl.Phys.B626 (2002)
363-376, hep-th/0110252

\bibitem{SLHypCh} B. L. Feigin, A. M. Semikhatov, V. A. Sirota, I. Yu
Tipunin, Nucl.Phys. B536 (1998) 617-656, hep-th/9805179 ;  B. L.
Feigin, A. M. Semikhatov, I. Yu Tipunin,  J.Math.Phys. 39 (1998)
3865-3905, hep-th/9701043

\bibitem{Yogendran} K.P Yogendran, JHEP 0501:036,2005, hep-th/0408114


\end{thebibliography}
\end{document}